\documentstyle[12pt,epsfig]{article}
\newcommand{\be}{\begin{equation}}  
\newcommand{\ee}{\end{equation}}  
\newcommand{\bear}{\begin{eqnarray}}  
\newcommand{\eear}{\end{eqnarray}}  
\newcommand{\ba}{\begin{array}}  
\newcommand{\ea}{\end{array}}

\newskip\humongous \humongous=0pt plus 1000pt minus 1000pt

\newif\ifdtup

\def\oldreffmt#1{\rlap{[#1]} \hbox to 2\parindent{}}

\def\figfmt#1{\rlap{Figure {#1}} \hbox to 1in{}}  
  
\def\ie{\hbox{\it i.e.}{}}	  
\def\eg{\hbox{\it e.g.}{}}	  
\def\etal{\hbox{\it et al.}}  
  
\def\beq{\begin{equation}}  
\def\eeq{\end{equation}}  
\def\bea{\begin{eqnarray}}  
\def\eea{\end{eqnarray}}  
\def\half{\frac{1}{2}}  
  
\def\bq{\begin{quote}}  
\def\eq{\end{quote}}

\def\half{\frac{1}{2}}       
\def \lta {\mathrel{\vcenter  
     {\hbox{$<$}\nointerlineskip\hbox{$\sim$}}}}  
\def \gta {\mathrel{\vcenter  
     {\hbox{$>$}\nointerlineskip\hbox{$\sim$}}}}   

\def \etal {{\it et al.}\ }  
\relax  

\newdimen\tdim  
\tdim=\unitlength

\begin{document}  
  \pagestyle{empty}  
\begin{titlepage}  
\def\thepage {}    
  \title{ \vspace*{1.5cm} \bf    
Geometrical Renormalization Groups:\\
Perfect Deconstruction Actions }   
\author{  
\bf  Christopher T. Hill$^1$ \\[2mm]   
{\small {\it $^1$Fermi National Accelerator Laboratory}}\\
{\small {\it P.O. Box 500, Batavia, Illinois 60510, USA}}
\thanks{e-mail: 
hill@fnal.gov }\\
}
\date{May, 2001}
\maketitle
\vspace*{-9.5cm}
\noindent
\begin{flushright}  
FERMILAB-Pub-03/057-T \\ [1mm]  
March, 2003  
\end{flushright}

\vspace*{10.0cm}  
\baselineskip=18pt  
  
\begin{abstract}  
{\normalsize By combining two distinct
renormalization group transformations, 
opposing  scale transformations, we obtain a composite transformation
which does not rescale the system, and drives it to a
``geometrical" fixed point, controlling the effective geometry
and locality.
The latticized (deconstructed) action for an
extra-dimensional field theory becomes  
a ``perfect action,'' with a linear ladder spectrum 
for $N$ modes. 
} 
\end{abstract}  
  
\vfill  
\end{titlepage}  
  
\baselineskip=18pt  
\renewcommand{\arraystretch}{1.5}
\pagestyle{plain}  
\setcounter{page}{1}

\section{Introduction}

When theories of
extra dimensions are latticized \cite{wang0}
or ``deconstructed" \cite{georgi}, 
we obtain gauge invariant effective Lagrangians for KK-modes
in $1+3$ dimensions. The structure of the compactified space is
mapped into the structure of the latticized matter theory.
The KK spectrum of the 
lattice action is not linear with only nearest
neighbor hopping terms, but takes the form
of a phonon spectrum,
\eg,
$m_n\sim \Lambda\sin(n\pi/N)$ for the modes on
a latticized compactified $S_1$. The spectrum is
only approximately linear for small  $n$, and 
the nearest neighbor latticized bulk
is not a true effective Lagrangian for 
$N$  KK modes.
Incorporating sufficiently many operators into 
${\cal{L}}_{eff}$ with only $N$ degrees of
freedom, we can always linearize the spectrum.

There is, however, a deeper issue here.
Deconstruction replaces the extra dimension by
a matter theory living in $3+1$ dimensions, and the
notion of the extra dimension's geometry is lost. 
What, therefore, is the symmetry
principle which governs the true ${\cal{L}}_{eff}$
for $N$ degrees of freedom with a linear ladder 
KK mode spectrum, intrinsic to the deconstructed theory
and making no reference to extra dimensions?  
This issue is surely related to the translational
invariance in the continuum extra dimensional
theory.  However,
if we write down an arbitrary continuum 
extra dimensional theory, we will have the
linear spectrum only if the theory has canonical
quadratic kinetic terms. Higher dimension 
operators, containing derivatives, must be absent
or suppressed. The linearity of the KK-mode spectrum has as much to
do with {\em locality} as with translational invariance.
We may therefore ask if there is a more general
statement about a theory which contains both the locality and geometry.

The set of all possible matter theories that 
contain the degrees of freedom of a given latticized
extra dimension, but which also contain more arbitrary
interactions, hopping terms, etc., has been dubbed ``theory space"
\cite{georgi}.
For our present discussion we
want the number of degrees of freedom, $N$, held
fixed, as a kind of ``microcanonical ensemble'' of theories.
We will argue that there is a particular 
operator, $Q$,  which acts upon the Lagrangians of the theory space,
mapping the theory space into itself.
$Q$ will be built out of the product
of pairs of renormalization group (RG)
transformations.  One of these RG transformations integrates
in more degrees of freedom (``decorates"
the theory) while the other RG transformation
thins the degrees of freedom by a block-spin
renormalization (Bell-Wilson transformation \cite{bell}).  

This contrasts typical RG transformations which
rescale a theory while thinning degrees of freedom, 
carrying a local ``reference scale of physics''
$\mu $ into a new scale $\lambda \mu$. 
$Q$ is a product of an ``up scaling" times a ``down scaling"
transformation, \ie,
$\lambda\times \lambda^{-1}$, and is net scale invariant.  
$Q$ is suited to the situation
of compactification of extra dimensions where there are two
fixed physical scales, 
the compactification radius $R$, and the fundamental,
or ``string''
scale, $\Lambda$ (which may be viewed as the perturbative
unitarity bound for longitudinal KK-mode scattering
in the deconstructed theory \cite{wang0}).  

The class of all $Q$'s and their fixed
points in theory spaces is no doubt very large and
rich. Recently we constructed a particular intriguing and
nontrivial example, a {\em fractal field theory of extra dimensions}
\cite{fractal}. 
The fractal theory is constructed on a lattice
with coordination number $s$ via a recursive
procedure in which $\sim s^{k+1}$
lattice sites are created at $k$th order. 
The $Q$ operation is defined by a pair of 
scaling operations which map $k\rightarrow k-1$,
but preserves scale, involving 
RG concepts borrowed from the Ising
model  \cite{Kramers,Onsager,Fish1}.
$Q$ becomes a symmetry operation  in
the ``continuum limit" $k\rightarrow \infty$, and 
one discovers a fixed point theory with
 a ``critical exponent,'' $\epsilon$, characterizing
the energy distribution of KK modes, ${\cal{N}}(E)
\sim (E/M)^\epsilon$.  Here $M$ is a nontrivial
RG invariant scale, the analogue
of $\Lambda_{QCD}$, and corresponds to the mass
of the first KK-mode, i.e. $M^{-1}$ acts like 
the effective compactification scale.
On scales above $M$ the theory 
effectively exists in $D+\epsilon$ dimensions,
where $D$ is integer, and
$\epsilon$ is an irrational number. 
The fractal theory is a nontrivial, 
candidate for a {\em finite quantum
field theory} to all orders of perturbation theory. The finiteness
traces to the RG transformation, implying that pointlike vertices
are meaningless, since they are replaced by simplices under the
transformation $Q$ (analogous to string theory having
undefined point-like vertices). 
Hence, $Q$ invariance can
define a nontrivial short-distance 
fractal theory with large distance Lorentzian geometry,
In this example $Q$ invariance is more fundamental 
in defining the theory than is geometry, which
becomes meaningless at short distances.
Perhaps the physical world does not respect
the notion of geometry at very short distances,
but, rather, is defined by recursion.

Thus, there exists a special point in
theory space, representing a Lagrangian ${\cal{L}}^*$,
which is a fixed point under $Q$, i.e.,
$Q({\cal{L}}^*) = {\cal{L}}^*$. 
$Q$ provides
an effectively local geometric theory, but makes no reference
to the existence of an extra dimension. 
$Q$ is defined only as a procedure that acts on theory space.

Let us first consider in general the relationship
of our problem, the construction of $Q$, 
for latticized extra dimension,
to the corresponding problem in an approximately scale invariant
theory, \eg, short distance lattice QCD.

\section{Perfect Actions}

The idea of constructing RG tranformations
and finding their fixed points to improve 
lattice actions lies in the arena of
``perfect actions'' from lattice QCD \cite{bell,wiese,hasen}.
One seeks improved lattice actions which, for QCD, are better
descriptions of the physics in the limit in which the 
dynamics is approximately scale invariant, \ie,
at very short distances where 
the stress tensor trace anomaly, 
is approaching zero,
$T_\mu^\mu \propto \beta(g)/g \rightarrow 0$.
The concept of perfect actions begins, 
in particular, with the block-spin renormalization group
transformation  of
Bell and Wilson \cite{bell}, and developed subsequently in QCD
lattice gauge theory, by  Wiese \cite{wiese}, 
Niedermayer, Hasenfratz, \etal, \cite{hasen}.
Perfect actions thus respect, in principle, a well
defined symmetry operation: they are invariant, or
fixed points, under block-spin renormalization group
transformations that are scale transformations. 
We say ``in principle'' because the implementation of
this idea is often only approximate, corrected perturbatively,
where possible. We evidently seek an 
analogue of the perfect action for a latticized compactified bulk,
and the principle which determines it.

A continuum free theory is trivially scale invariant under Bell-Wilson
BSRG transformations when a certain parameter of
the transformation, $\lambda$, is
chosen to drive the theory toward the Gaussian Fixed point.  
With a less trivial dispersion
relation, $p_0^2 = \vec{p}^2 + \vec{p}^4/M^2 + ...$ the theory
flows into the infrared free field theory $p_0^2 = \vec{p}^2$
under successive applications of the
Bell-Wilson transformation for $\lambda^*=1/\sqrt{2}$ (Bell and Wilson
introduce the parameter $b$ which corresponds to our
$\lambda$ up to a dimension dependent factor; presently $d=1$ 
in their language, and the critical
value of $\lambda=\lambda^*$ matches $b^*$).  
Perfect actions for QCD are in principle the fixed point actions
under this transformation. They are, in principle,
implemented in the fully interacting
quantum field theory.  They are typically quasilocal, involving
suppressed (next-to)$^n$ nearest neighbor interactions.  
Approximate 
forms have been  constructed, \cite{hasen},
and the Gaussian  perfect action  
gives the best starting point for
a perturbative quantum perfect action. 

In the case of compactified extra dimensions and
their deconstructed description,  the situation
with respect to scale invariance is drastically
different than that of QCD. Here
we {\em do not}
have scale invariance. We have instead
two relevant scales: (1) a high energy or fundamental, or ``string
scale," $\Lambda$, and (2) the low energy 
compactification scale $M = 1/R$.
The compactification 
scale $\sim 1/R$  is identified with the mass of the lowest
KK mode.  RG transformations of interest must therefore
act within the range of scales $\Lambda >> \mu >> M$, but
must not affect $\Lambda$ and $M$, 
which define the theory. 
Indeed, a continuum extra dimensional theory is generally
never scale invariant. The stress
tensor trace is nonzero in $D\neq 4$, and the coupling constant
is dimensional. This is the origin of
the power-law running of the coupling constant.
In deconstruction, the corresponding $D=4$
stress tensor trace matches the $D\neq 4$ stress
tensor trace through the presence of the ``linking
Higgs'' fields, which have nonzero VEV's of order $\Lambda$
explicitly breaking scale invariance.

In deconstruction, the continuous
translational symmetry in an
extra dimension is replaced by a discrete
$Z_N$ symmetry. The kinetic
terms become
$Z_N$ invariant hopping terms. Locality
implies that hopping terms do not have strong linking to
distant sites. This is the key to having
anything resembling a ladder spectrum. For example, if one constructs a
``complete'' lattice of $N$ sites, 
linking each site to every other site with a hoping
term strength $\Lambda$, as
in Figure (1),
the spectrum will be have a single zero mode, and $N-1$
degenerate levels of mass $\sim \sqrt{N}\Lambda$. This is as
far from a ladder spectrum as one can get.

\begin{figure}[t]    
\vspace{4cm}    
\includegraphics{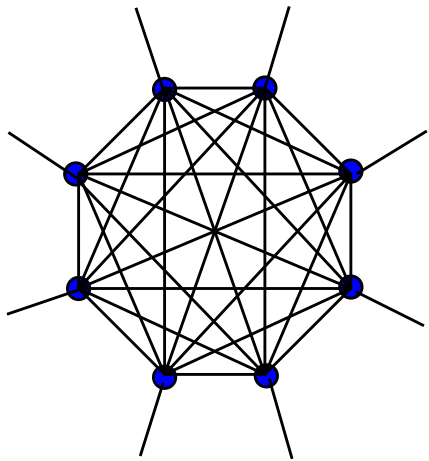}    
\vspace{1.5cm}    
\caption{\small A ``complete octagon'' has each
of the 8 sites linked to every other site. The eigenmode
spectrum is one zero mode and 7 modes of mass $\sim \sqrt{8}\Lambda$.
}   
\label{third}     
\end{figure}

It is locality, therefore, together with $Z_N$
invariance, that 
sets up the hierarchy between the high scale
$\Lambda$ and the compactification scale
$1/R$, and uniformly
populates the spectrum on scales $\Lambda > \mu > 1/R$ with KK modes.
 Nevertheless, any restriction to local, 
mainly nearest neighbor, links appears to be 
arbitrary
from a ``theory space'' point of view.
The $Q$ symmetry however, will select 
the fixed point theory, enforcing
quasi-locality and a linear ladder
spectrum.

We can understand $Q$ with the following
metaphor. Suppose we are climbing a tall 
cell-phone tower ladder, and
we are in the middle of the ladder
at a scale $\mu$, roughly
midway between the top of the ladder ($\Lambda$) and 
bottom  ($M$). When we are midway
up the ladder ($\Lambda>> \mu >> M$),
we are far from any fundamental
scale, and the
effective Lagrangian should be insensitive to changes in $\mu$. 
The ladder 
reflects an approximate symmetry in momentum space
in that, ``hopping up'' several rungs, the ladder 
should appear to us to be the same physical environment. 
Suppose we double $\Lambda/M$, thus doubling the height
of the ladder while holding the
rung spacing on the ladder fixed. 
This maps our ladder with $N$
rungs into one with $2N$ rungs. 
Now we follow this 
with another operation that integrates out $1/2$
of the rungs and rescales the ladder height
by $1/2$. Under these combined
operations we have no net rescaling,
but trivially recover the original 
linear ladder with $N$ rungs.

More generally, however,
if the ladder is not uniform, we might double the
number of rungs by some procedure,
$k$ times, thus increasing its height by $2^k$. Then  we
``integrate out'' half the rungs $k$ times by averaging
over the locations of rungs that are integrated, suitably smearing
out the local inhomogeneities. We expect then that
an arbitrary inhomogeneous ladder becomes flat in the large $k$ limit.
Thus, while the spectrum of the nearest neighbor theory is
not flat, having the phonon $m_n\sim \Lambda\sin(n\pi/N)$ structure,
under application of this $k$th order RG
transformation we expect it to flow
toward a fixed point, until it flattens, $m_n\sim \Lambda n\pi/N$.
An action which is a fixed point under
this transformation should be the perfect action we seek.

In Section 3 we will make this more precise in the language
of ``decoration'' and ``dedecoration'' transformations,
which are defined in configuration space. We first
define the product of a decoration
transformation $D$ and a dedecoration transformation $D^{-1}$ 
to be the identity transformation. We illustrate
our general technique by
carrying out an analysis of the trivial
case of $k$ decorations followed by
$k$ dedecorations, $(D^{-1})^kD^k$ 
to set up the subsequent nontrivial case of interest. 

We then consider decoration transformations
followed by Bell-Wilson (BW) transformations
\cite{bell} acting on scalar field theories.
The BW transformation acts in momentum space
to define the block spin variables, and introduces
a new parameter, $\lambda$, which is the ``wave-function 
renormalization constant'' of the block-spin variable.
Thus, we perform a decoration transformation $D^k$
followed by a Bell-Wilson transformation $k$ times, $B^k$,
which groups the original
sites into ``block spins'' and maps the theory back to 
the original $N$
branes.  We thus have $Q$ defined through
$Q^k = B^kD^k$ and we are interested in the formal
limit  as $k\rightarrow \infty$.

The Gaussian fixed point of the
BW transformation is found for the special 
value $\lambda=\lambda^*=1/\sqrt{2}$.
The spectrum, under 
the $Q$ transformation for large $k$
flows toward the linear ladder spectrum.
Indeed, the spectrum has the form
$m_n^{(k)} \rightarrow 2^{k+1} \Lambda\sin(2^{-k}n\pi/N) $ 
under $B^kD^k$, 
and flows toward the flat
$m_n^{(k)} \rightarrow 2\Lambda n\pi/N$ spectrum
as $k\rightarrow\infty$.
The action of the deconstructed theory which
is itself a fixed point under $B^kD^k$ as $k\rightarrow \infty$  
corresponds to ``a perfect action,'' analogous to
perfect actions in lattice QCD.
This codifes the sense in
which the ladder spectrum is special,
within the context of theory space.


\section{Transformations for
Deconstructed Lattice Scalars in $1+4$}

We begin by considering transformations
which augment or thin the degrees of freedom
of  $1+3$ theories of many complex scalar fields.
These transformations stem from 
symmetries noticed long ago in the Ising model, 
\cite{Kramers,Onsager,Fish1}.   
In the language of
Ising models a single spin$_1$-link-spin$_2$ combination 
in the Hamiltonian can
always be ``decorated,'' \ie, 
written as a spin$_1$-link-spin${}'$-link-spin$_2$ interaction.
That is, we can ``integrate in'' the new spin${}'$,
or ``decorate'' the original single link. Thus, an $N$-spin
system can be viewed as a $2N$ spin system upon decorating.
The decorations can be arbitrarily complicated, involving many
new spins. Conversely, we can ``integrate out'' or ``dedecorate''
the spins internal to a chain whose endpoint spins are then renormalized
(Fig.(2)).
   
\begin{figure}[t]    
\vspace{4cm}    
\includegraphics{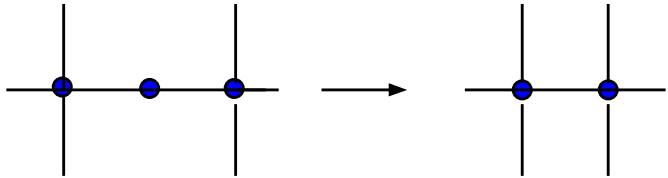}    
\vspace{1.5cm}    
\caption{\small The $3$-chain $\rightarrow$ $2$-chain dedecoration 
transformation integrates out the internal field and renormalizes the 
endpoint fields' kinetic terms and mass terms.}   
\label{first}     
\end{figure}

Decoration is an exact transformation for Ising spins, and
continuous spins (e.g., ``spherical models'' are spin systems
which correspond to our models in the absence of kinetic terms). 
Presently our ``spins'' are fields that 
have $1+3$ kinetic terms. For us decoration and 
dedecoration
transformations are 
exact transformations only in the limit
of very large cut-off $\Lambda$.  This happens because we perform
decoration transformations truncating on
quartic derivatives, such as $|\partial^2\phi_a|^2/\Lambda^2$.
This is, nonetheless,  a good approximate 
transformation in the $\Lambda\rightarrow \infty$
limit, or for the low lying states in the spectrum. 
These transformations become
symmetries in the continnum
limit
when the theory is classically scale
invariant, \ie, $\mu^2=0$ and $\Lambda\rightarrow \infty$.   
The  
$1+3$ kinetic terms undergo renormalizations
under these transformations, and 
thus distinguish the present construction
from that of a spin model.

\subsection{Decoration 
Transformations}

Consider  an $N$ complex scalar field Lagrangian in $1+3$,
which can be viewed as a deconstructed  $S_1$ compactified extra
dimension  with periodic boundary conditions:
\be
{\cal{L}}_N = Z_N\sum^{N}_{a=1} |\partial \phi_a|^2 
- \Lambda_N^2 \sum^{N}_{a=1} |\phi_a -\phi_{a+1}|^2 
- \mu_N^2 \sum^{N}_{a=1} |\phi_a|^2 
\label{one}
\ee
and assume periodicity, hence
$\phi_{N+a} = \phi_a$. It is convenient to
allow for noncanonical normalization of the kinetic
terms, and we thus display the arbitrary wave-function renormalization
constant $Z_N$.

A decoration transformation replaces ${\cal{L}}_N$
by a new Lagrangian  ${\cal{L}}_{2N}$ with $2N$
fields and new parameters.
\be
{\cal{L}}_N \rightarrow {\cal{L}}_{2N} = Z_{2N}\sum^{2N}_{a=1} |\partial \phi_a|^2 
- \Lambda_{2N}^2 \sum^{2N}_{a=1} |\phi_a -\phi_{a+1}|^2 
- \mu_{2N}^2 \sum^{2N}_{a=1} |\phi_a|^2 
\label{one2}
\ee
and again assume periodicity, hence
$\phi_{2N+a} = \phi_a$
The new parameters are chosen so that upon integrating out 
every other field (e.g., those with even $a$ in the sums), we
recover the original Lagrangian ${\cal{L}}_N $. 

We define the new decorated Lagrangian
parameters by demanding 
that ${\cal{L}}_{2N}$ be equivalent to
${\cal{L}}_N$ upon ``dedecorating,'' \ie, integrating out
every other field.  We seek
the relation between the parameters $X_{2N}$ of
${\cal{L}}_{2N}$, and the $X_{N}$ of
${\cal{L}}_{N}$.
It is useful to consider
${\cal{L}}_{2N}$ as a sum over 3-chains:
\be
{\cal{L}}_{2N} = \sum^{2N-1}_{n\;odd} {\cal{L}}_{n,n+2} 
\label{one2}
\ee
Each 3-chain involves three fields. 
The first 3-chain is:
\be
{\cal{L}}_{1,3} = \half Z_{2N}
(|\partial \phi_{1}|^2 + 2|\partial \phi_{2}|^2
+|\partial \phi_{3}|^2)
- \Lambda_{2N}^2|\phi_1 -\phi_{2}|^2 - \Lambda_{2N}^2|\phi_2 -\phi_{3}|^2 
- \half \mu_{2N}^2 (|\phi_{1}|^2+2|\phi_{2}|^2+|\phi_{3}|^2)
\label{first}
\ee
The fields $\phi_1$ and $\phi_3$ share
half their kinetic terms and $\mu^2$ terms
with the adjacent chains, thus carry the normalization factors
of $1/2$ within the chain.
$\phi_2$ can be thus viewed as a ``decoration'' of the
chain. We integrate out the internal field $\phi_2$ and obtain
an equivalent renormalized chain.
Integrating out  $\phi_2$:
\bea
{\cal{L}}_{1,3} & = & \half Z_{2N}(|\partial \phi_1|^2 + |\partial \phi_3|^2) - 
(\Lambda_{2N}^2+\half \mu_{2N}^2)(|\phi_1|^2+|\phi_3|^2)
+ 
\nonumber \\
& & + \Lambda_{2N}^4 (\phi_1+\phi_{3})^\dagger \frac{1}{Z_{2N}\partial^2 +
2\Lambda_{2N}^2+\mu_{2N}^2}(\phi_1+\phi_{3})
\label{first2}
\eea
Expanding in the derivatives to ${\cal{O}}(\partial^2)$,
integrating by parts, regrouping terms, and
relabeling parameters gives:
\be
{\cal{L}}_{1,3} = \half Z_N(|\partial \phi_1|^2 + |\partial \phi_3|^2)- 
\Lambda_{N}^2|\phi_1 -\phi_{3}|^2 -\half\mu_{N}^2
(|\phi_1|^2 + |\phi_3|^2)
-\delta_N|\partial (\phi_1-\phi_{3})|^2 + {\cal{O}}(\partial^4/\Lambda^2)
\label{first3}
\ee
where we now have the relationship
between the  parameters of
eq.(\ref{first2}) and eq.(\ref{first3}):
\bea
Z_N & = & Z_{2N}\frac{8\Lambda_{2N}^4 + 4\Lambda_{2N}^2\mu_{2N}^2 
+\mu_{2N}^4}{4\Lambda_{2N}^4 + 4\Lambda_{2N}^2\mu^2 +\mu_{2N}^4}
\approx 2Z_{2N}
\nonumber
\\
\Lambda_{N}^2 & = & \frac{\Lambda_{2N}^4}{2\Lambda_{2N}^2 +\mu_{2N}^2}  \approx
\half\Lambda_{2N}^2
\nonumber
\\
\mu_{N}^2 & = & \frac{4\mu_{2N}^2 \Lambda_{2N}^2+\mu_{2N}^4}{2\Lambda_{2N}^2
 +\mu_{2N}^2} 
\approx 2\mu_{2N}^2
\nonumber
\\
\delta_N & = & \frac{Z_{2N}\Lambda_{2N}^4}{(2\Lambda_{2N}^2 +\mu_{2N}^2)^2}\approx
\frac{1}{4}Z_{2N}
\label{renorms}
\eea
Finally,
we relabel the indices so they are sequential, i.e., $a=2b-1$ where
$b$ runs from $(1, N)$, and $\phi_{2b-1} \equiv \phi'_b$.
We then rewrite the Lagrangian:
\be
{\cal{L}}_{2N} =  Z_{N}\sum^{N}_{b=1} |\partial \phi'_a|^2 
- \Lambda_{N}^2 \sum^{N}_{b=1} |\phi'_b -\phi'_{b+1}|^2 
- \mu_{N}^2 \sum^{N}_{b=1} |\phi'_b|^2 
-\delta_N\sum^{N}_{b=1}|\partial (\phi'_b-\phi'_{b+1})|^2 
+ {\cal{O}}(\partial^4/\Lambda^2)
\label{six}
\ee
Thus, we have returned to the original theory, ${\cal{L}}_{N}$,
modulo the $\delta_N$ term and the higher derivative terms.
The presence of the $\delta_N $ term, and higher
derivatives, reflects the fact
that our dedecoration transformation is not exact. These terms
are ``irrelevant operators,'' however.
The $\delta_N $ term has been written in the indicated form
because, though it superficially appears to be a relevant $d=4$
operator, 
it too is a quartic derivative on the lattice, \ie, ($\partial^2$ in $1+3$)
$\times$(a nearest neighbor hopping term on the lattice). 
It effects
only the high mass limit of the KK mode spectrum. 
It will therefore be dropped
for consistency with the expansion 
to order $\partial^4/\Lambda^2$.

In eqs.(\ref{renorms}) we have written the approximate forms of the
renormalizations of the parameters in the large
$\Lambda_{2N}$ limit.
Note that the $\mu^2$ term is multiplicatively renormalized.
This owes to the fact that it is the true scale-breaking term in the
theory when the lattice is taken very fine, and $\Lambda$ terms 
become  derivatives, \ie, as $\mu\rightarrow 0$ the theory
has a zero-mode. Since it alone breaks the symmetry
of scale-invariance, elevating the zero-mode,
it is therefore multiplicatively renormalized in free field theory. 

We now attempt to define $Q=(D^{-1})^kD^k$  as a product
of $k$ decorations followed by $k$ dedecoration
transformations. This will, of course, be
the identity, and we'll recover the original
Lagrangian, but it illustrates the technique
for the subsequent nontrivial
case involving the BW transformation. We first 
iterate the decoration transformation of eq.(\ref{one})
$k$ times to obtain a $2^kN$ decorated theory:
\be
{\cal{L}}_{2^kN} = 
Z_{2^kN}\sum^{2^kN}_{a=1} |\partial \phi_a|^2 
- \Lambda_{2^kN}^2 \sum^{2^kN}_{a=1} |\phi_a -\phi_{a+1}|^2 
- \mu_{2^kN}^2 \sum^{2^kN}_{a=1} |\phi_a|^2 
\label{one11}
\ee
and again assume periodicity, hence
$\phi_{2^kN+a} = \phi_a$.
The new parameters are chosen so that upon $k$
applications of the dedecoration
transformation, we
recover the original Lagrangian ${\cal{L}}_N $.

We diagonalize eq.(\ref{one11}) with periodic
compactification:
\be
\phi_a = \frac{1}{\sqrt{2^kN}}\sum^{2^kN}_{n=1} e^{2\pi i na/2^kN} \chi_n;
\qquad \phi_{a + 2^kN} = \phi_a
\qquad \makebox{note:}\;\;S = \sum^{2^kN}_{a=1} e^{2\pi i(n-m) a/2^kN} 
= 2^kN\delta_{nm}
\ee
whence:
\be
{\cal{L}}_{2^kN}(q^2) = Z_{2^kN}\sum^{2^kN}_{n=1} |\partial \chi^{(k)}_n|^2 -  
\sum^{2^kN}_{n=1} (4\Lambda_{2^k N}^2\sin^2(\pi n/2^kN) +\mu_{2^kN}^2)|\chi^{(k)}_n|^2
\label{two11}
\ee
Note that 
we have the zero-mode, corresponding to
$n=0$,  which is a singlet 
(equivalent to choosing $n=2^kN$, which
is outside the first Brillouin zone). 
Each level within
the first Brillouin zone with $n\neq 0$ is degenerate with
another level $ 2^kN-n$, thus forming a doublet. This doubling of 
energy levels is physical, corresponding
to the mode expansion in $x^5$ in terms of $1$, $\sin k_nx^5$
and $\cos k_nx^5$, where the sine and cosine terms are
degenerate modes (or equivalently, left--movers and right--movers).
With orbifold compactification each level would be a singlet, and
what we say presently works as well in the orbifold case.

For KK modes of
momentum $q^2$, we have
from the momentum space form
of eq.(\ref{two11}), 
\be
\label{compact}
{\cal{L}}_{2^kN} = \sum_q \sum^{2^kN}_{n=1} \omega_n^{(k)}(q^2) 
|\chi^{(k)}_{q,n}|^2 
\qquad
\omega_n^{(k)}(q^2) = Z_{2^kN}q^2 
- 4\Lambda_{2^kN}^2\sin^2(\pi n/2^{k}N) - \mu^2_{2^kN}
\ee
The parameters of the $k$-th order decorated theory
may be written in terms of the parameters of the 
${\cal{L}_N}$ theory from eqs.(\ref{renorms}):
\beq
Z_{2^kN}   =  2^{-k}Z_N 
\qquad
\Lambda_{2^kN}^2  =  2^{k}\Lambda_{N}^2
\qquad
\mu_{2^kN}^2 = 2^{-k}\mu_{N}^2 
\label{renorms2}
\eeq
Note that, while we're writing
things in momentum space, the actual dedecorations
are done in configuration space, \ie, we return to configuration
space to integrate out
alternating fields. At the $\ell$th iteration
the number
of fields remaining is $2^{k-\ell}N$ and
we replace the current $\chi^{(\ell)}_{q,n}$
coefficents by new set of $\chi^{(\ell-1)}_{q,n}$, with
$2^{k-\ell-1}N$ Fourier coefficients.  We then rediagonalize  
to obtain the new momentum space expression for
$\omega_n^{(\ell)}$. This replaces the previous
value of $2^{k-\ell} N$ by $2^{k-\ell-1} N$ in the argument of
the $\sin^2(\pi n/2^{x}N)$ of eq.(\ref{compact}).

We thus obtain $(D^{-1})^kD^k$,  from the rescalings
of eq.(\ref{renorms}), \ie,
$\omega_n^{(k)}\rightarrow \omega_n^{(0)}$, where:
\beq
\omega_n^{(0)} = 2^{-k}Z_N q^2 
- 2^{-k+2}\Lambda_{N}^2\sin^2(\pi n/N) - 2^{-k}\mu_{N}^2
\eeq
where $2^0 N = N$ now appears in the argument of
$\sin^2(\pi n/N)$.

Renormalizing the fields 
to canonical normalization  leads to: 
\beq
\omega_n^{(0)} \rightarrow  q^2 
- 4\Lambda^2\sin^2(\pi n/N) - \mu^2
\eeq
where:
\beq
\Lambda^2\equiv \Lambda^2_N/Z_N \qquad  \mu^2\equiv \mu^2_N/Z_N
\eeq
We have recovered the original
theory.  Decoration and dedecoration transformations
are inverses of one another. Indeed, that is how they were
constructed. 

As we now see, to define $Q$, such that the 
compactification scale $1/R = M= \Lambda\pi/N$ and
number of modes $N$ is held fixed, 
such that we obtain a linear ladder spectrum,
we must follow decoration
by a transformation that thins degrees of freedom
in momentum space.

\subsection{Bell-Wilson Transformation}

We now consider $Q^k = B^kD^k$
where is $B^k$ a degree-of-freedom thinning transformation 
that  acts upon the decorated theory,
$D^k({\cal{L}}_{N}) = {\cal{L}}_{2^k N}$ of eq.(\ref{one11}). $B$
is the BW transformation, 
which, contrary to dedecoration,
is defined  in momentum space. Moreover,
the rescaling from $\ell\rightarrow \ell-1$ is now
controlled by a new parameter, $\lambda$,
which is associated with the block-spin 
wave-function normalization. 
For a special
choice of $\lambda=\lambda^*=1/\sqrt{2}$
we will scale toward a Gaussian fixed point theory
which recovers the linear ladder spectrum. 

The parameters of ${\cal{L}}_{2^k N}$, the $X_{2^k N}$,
are given in terms of $X_{N}$ by eq.(\ref{renorms2}).
We again diagonalize eq.(\ref{one11}) with periodic
compactification obtaining eq.(\ref{two11}) and
adopt the compact notation
for ${\cal{L}}_{2^k N}$ of eq.(\ref{compact}) in momentum space
where the field's $2^kN$ Fourier coefficients
are denoted $\chi_{q,n}^{(k)}$.

The Bell-Wilson transformation can be
phrased as a mapping from the action ${S}_{(2^kN)}(\chi^{(k)})$ 
as a functional of the ``old'' $2^k N$ variables, $\chi^{(k)}_n$ 
to a new actiuon, ${S}_{(2^{k-1}N)}(\chi^{(k-1)})$ in the ``new''
$2^{k-1} N$ variables, $\chi^{(k-1)}_n$:
\be
e^{i{S}_{(2^{k-1}N)}(\chi^{(k-1)})} 
= \int D\chi^{(k)} \; T(\chi^{(k-1)}, \chi^{(k)}) 
e^{i{S}_{(2^{k}N)}(\chi^{(k)})}
\ee
We choose a Gaussian block spin redefinition
in momentum space, of the form:
\be
\label{bw}
 T(\chi^{(k-1)}, \chi^k) = \exp\left(i K \sum_{n=0}^{2^{k-1}N} |\chi^{(k-1)}_n 
 -\lambda\chi^{(k)}_n|^2  \right)
\ee
Here $K$ is a (large) mass${}^2$ scale parameter, and is formally
arbitrarily defined relative to the normalizations
of the $\chi$'s. $K$ is introduced to engineer a functional
delta-function, locking the new
$\chi^{(k-1)}_n$ and the old $\chi^{(k)}_n$
together. $K$ should exceed the largest physical scale
in the problem, which is the hopping parameter $\Lambda_{2^k N}^2$.
Let us therefore define:
\beq
K = F \Lambda_{2^k N}^2 \qquad  F>>1.
\eeq
The key to the BW transformation is that we
we integrate out half of
the old momentum space modes for a given value of $\lambda$.
Half of the $\chi^{(k)}_n$ modes, from $n=1$ to $n=2^{k-1}N$, 
are mapped into the new 
$\chi^{(k-1)}_{n}$ modes,
while the remaining high momentum, or short distance,
$n> 2^{k-1}N$ modes are simply  integrated out. 
The $\chi^{(k-1)}_{n}$ modes are
labelled by indices that run from $j=1$ to $2^{k-1}N$.
The fact that only half of the $\chi^{(k)}_n$ 
momentum modes are mapped
into the transformed action means that the $\chi^{(k-1)}_n$ modes
are ``block spins'' in configuration space.
It is not hard to see, in configuration space, that
one BW transformation yields  $2^{k-1}N$ new fields, $\phi_a^{new}$,
written as blocks spins of the  $2^{k}N$ old fields, $\phi_a^{old}$
as:
\beq
\phi_a^{new} 
= \frac{\lambda}{2\pi} \sum_{b=1}^{2^kN} \frac{|\sin(\pi(a-b/2))|}{|a-b/2|}\phi_b^{old}
\eeq
where we have labelled the $2^{k-1}N$ new
fields as $1\lta a\lta 2^{k-1} N$, and the old
ones as $1\lta b\lta 2^{k} N$, and we have 
absorbed phases into $\phi_b^{old}$.
 $\lambda$ plays the fundamental role of 
defining the wave-function
normalization of the new spin block
spin variables. 

Successive application
of the BW transformation $k$ times leads to:
\be
e^{iS^{(0)}(\chi^{(0)})}
=\int D\chi^{(k)}..D\chi^{(1)}
e^{iK\sum_{n=1}^{2^{k-1}N}|\chi^{(k)} - \lambda\chi^{(k-1)} |^2}
...e^{iK\sum_{n=1}^{2N}|\chi^{(1)} - \lambda\chi^{(0)} |^2}
e^{- \sum_{n=1}^{2^kN}\omega^{(2^kN)}_n |\chi_n^{(k)} |^2}
\ee
Under the $\ell$th transformation of the $\ell$ variables
into the
$\ell-1$ variables we find:
\be
S_{(2^{(\ell-1)}N)}(\chi^{(\ell-1)}) = K \sum_q \sum_{n=1}^{2^{(\ell-1)}N}
\frac{\omega_n^{(\ell)}(q^2)}{K\lambda^2 +
\omega^{(\ell)}_n(q^2)} |\chi_{q,n}^{(\ell)}|^2
\ee
hence:
\be
\omega_n^{(\ell-1)} = \frac{\omega_n^{(\ell)}}{(\lambda^2)
+\omega_n^{(\ell)}/K} 
\ee
This is a recursion relation which 
we can easily
solve for the general map from $\omega^{(k)}$,
corresponding to ${\cal{L}}_{2^k N}$, 
back to $\omega^{(0)}$, corresponding to ${\cal{L}_{N}}$.
We obtain:
\be
S_{N}(\chi^{(0)}) =  \sum_q \sum_{n=1}^{N}\omega_n^{(0)}(q^2)|\chi_{q,n}^{(0)}|^2
\ee
where:
\be
\omega_n^{(0)} = \frac{\omega_n^{(k)}}{(\lambda^{2k}) 
+c_k\omega_n^{(k)}/K} 
\qquad c_k = {\lambda^{-2k}}\left( \lambda^{2+2k}-1\right)\left(\lambda^2-1 \right)^{-1}
\ee
This is the general renormalization group for the kinetic
term part of the theory. 

For KK modes of
momentum $q^2$, we have for the $k$th order
decorated theory, 
\be
\omega^{(k)}(q^2) = Z_{2^kN}q^2 
- 4\Lambda_{2^kN}^2\sin^2(\pi n/2^{k}N) - \mu^2_{2^kN}
\ee
From the original decoration transformation
acting on ${\cal{L}}_N$ we have: 
\beq
Z_{2^kN}\rightarrow 2^{-k}Z_{N}
\qquad 
\Lambda_N^2 = Z_{N}\Lambda^2
\qquad
 \Lambda_{2^kN}^2 = 2^{k}Z_{N}\Lambda^2 
\qquad  
K = 2^{k} F Z_{N}\Lambda^2
\eeq
through the RG results of eq.(\ref{renorms2}).
The scale $M =\pi\Lambda_N/N $ is the RG invariant 
mass of the first KK-mode.
We also have:
\beq
 Z^{-1}_{2^kN}\mu^2_{2^kN} =Z^{-1}_{N}\mu_{N}^2 \equiv  \mu^2 
 \qquad\makebox{hence:}\qquad
 \mu^2_{2^kN} = 2^{-k}\mu^2 
\eeq
where $\mu^2$ is the RG invariant physical mass scale.

We note that a key difference between the
BW transformation and the dedecoration transformation, is
that in the $\omega^{\ell}_n$ at the $k-\ell$th
order of iterating the transformation, the 
the KK-mode mass term involves $\sin^2(\pi n/2^kN)$, and
{\em not} $\sin^2(\pi n/2^\ell N)$!  
This is a direct consequence
of the fact that the BW transformation is defined through eq.(\ref{bw})
{\em in momentum space}, and we do not return to configuration space
to integrate out fields!

Finally, for large $k$ and we see that
\beq
c_k \rightarrow \frac{1}{1-\lambda^2}\lambda^{-2k}\qquad \lambda < 1;
\qquad
c_k\rightarrow 1 \qquad \lambda > 1.
\eeq
The final action after $k$ iterations
of the BW transformation is summed over
$N$ degrees of freedom,
and we can therefore write, for $\lambda < 1$:
\be
\omega^{(0)}_n 
=  \sum_{n=1}^{N}\frac{2^{-k}Z_{N}\left(q^2 
- 2^{2k+2}\Lambda^2\sin^2(\pi n/2^kN)- \mu^2\right)|\chi_n^{(0)}|^2}{
\lambda^{2k}+
\lambda^{-2k}2^{-2k}Z_{N}
\left[(q^2-2^{2k+2}\Lambda^2\sin^2(\pi n/2^kN)-\mu^2)
/(1-\lambda^2)Z_N F \Lambda^2\right]
}
\ee
The generality of the
BW transformation
is contained in the freedom to choose the 
parameter $\lambda$. If $\lambda  
< 1/\sqrt{2}$ we see that the second term
in the denominator dominates in the large $k$ limit
and the theory becomes a static (nonpropagating) system,
\beq
\omega^{(0)}_n(q^2)
\rightarrow   \sum_{n=1}^{N}{\lambda^{2k}2^{k}(1-\lambda^2)Z_N F \Lambda^2
|\chi_n^{(0)}|^2}
\eeq
This is essentially an auxilliary field theory 
with the effective
kinetic term normalization set to zero.

For the special case $\lambda = \lambda^* 
\equiv 1/\sqrt{2}$
we have $c_k \rightarrow (4/3)\lambda^{-2k}$
and the theory approaches a nontrivial fixed point,
called the ``Gaussian fixed point.'' We can exercise
our freedom to define
the original kinetic term normalization constant $Z_N=1$. 
For $\lambda=\lambda^* $ we have:
\bea		       
\omega^{(0)}_n  & = &
\frac{(q^2-2^{-2k+2}\Lambda^2\sin^2( \pi n/2^k N)-\mu^2)}
{1+4(q^2-2^{-2k+2}\Lambda^2\sin^2( \pi n/2^k N)-\mu^2)/3F \Lambda^2}
\\ \nonumber 
& \approx&  q^2-4\Lambda^2 n^2\pi^2/N^2 - \mu^2 +
{\cal{O}}(1/F).
\eea
which leads to the ``linear ladder spectrum'' of $N$ KK modes
as $k\rightarrow \infty$:
\be
m_n^2 = 4\Lambda^2(\pi n/N)^2+\mu^2
\ee
 Therefore, $Q$ defined as a $k$th order decoration
of an $N$-mode containing deconstructed theory,
followed by the $k$th order Bell-Wilson transformation
at the fixed point $\lambda = \lambda^* 
\equiv 1/\sqrt{2}$ yields a theory with $N$-modes
and the linearized spectrum.

If $\lambda  
> 1/\sqrt{2}$ we see that the first term
in the denominator dominates in the large $k$ limit
and the theory scales as:
\be		       
\omega^{(0)}_n \rightarrow 
\lambda^{-2k}2^{-k} Z_{N}(q^2-2^{2k+2}\Lambda^2\sin^2( \pi n/2^k N)- \mu^2)
\ee
We can  define
$\lambda^{-2k}2^{-k} Z_{N} =\eta^{-k} $ where $\eta > 1$:
\be		       
\omega^{(0)}_n \rightarrow 
\eta^{-k}(q^2- 4\Lambda^2( \pi n/N)^2-\mu^2)
\ee
Therefore,  $\lambda  
\gta 1/\sqrt{2}$ defines a fixed
line of theories with ladder 
spectrum.  The Gaussian fixed point
is presumably stable in the presence of interactions,
as in the nonlinear generalization of Bell and Wilson.

\section{Discussion and Conclusion}  

We have considered RG transformations that
act in a theory space of $N$ degrees of freedom.
This theory space {\em a priori} knows nothing
of extra dimensions and geometry. Nonetheless,
a particular RG transformation $Q$, constructed
from a net scale-invariant pair of RG transformations,
has  a fixed point corresponding to an effective theory 
of a compactified geometrical extra dimension. 

More specifically, we have observed that
the combination of a $k$-fold decoration
transformation, followed by a $k$-fold 
Bell-Wilson transformation
can produce a theory with (i) no rescaling of the number of
degrees of freedom, $N$, and (ii) a fixed point
with a linear KK-mode spectrum, when 
$Q$ is defined with its parameter $\lambda = \lambda^* 
\equiv 1/\sqrt{2}$. This
corresponds to the Gaussian fixed point 
of the Bell-Wilson transformation. 

This has been a preliminary look
at the problem of constructing more
general $Q$ operators. There remains a great
to do. It
should be possible to implement this transformation
in an interacting theory, in analogy to QCD \cite{hasen}. 

$Q$ as we've defined
it requires that we exit the microcanonical ensemble
of $N$ modes, passing to $2^kN$ modes, as an intermediate step.
This is fine for free field theory, but apparently
undesirable in the interacting case because 
the physical interacting
theory is bounded in $N$ by
the perturbative unitarity.  
This latter issue
however, may be a red-herring.
It may be possible to make the step outside
the microcanonical ensemble 
if we also make the effective coupling
scale, \eg, $2^{-k}$ times smaller upon
decoration, which defines the
decoration transformation as holding the 
 high energy coupling constant at the scale 
 $\Lambda_{2^kN}$ fixed.

It would be useful to connect to the differential
form of the RG, which is more
familiar in QFT perturbation theory.  Ultimately, we 
would prefer a more
abstract definition of $Q$ that lives
entirely within the microcanonical
ensemble of fixed $N$ theories in theory space.
A possible route to a more general $Q$
may involve abstracting from the present procedure.
It is a simple matter to write the fixed
point theory as a Lagrangian of the form:
\be
{\cal{L}}^*_N = \sum^{N}_{a=1} |\partial \phi_a|^2 
- \Lambda_N^2 \sum^{N}_{a=1} \sum^{N}_{b=1}c_b|\phi_a -\phi_{a+b}|^2 
- \mu_N^2 \sum^{N}_{a=1} |\phi_a|^2 
\label{final}
\ee
where the $c_b$ are chosen to yield the flat spectrum.
This involves a simple Fourier transform
of the flat spectrum.
It would be instructive to frame the action of 
the $k$th order $Q^k$ on
a generic ${\cal{L}}'_N $, containing arbitrary coefficients,
$c_b'$, as a mapping of  $c_b' \rightarrow c_b$,
and to observe the approach to the fixed point theory.
This may permit an abstraction of $Q$,
which discards the decoration and BW transformations
altogether, and a $Q$ that is
more practically useful than the detailed procedure we
have outline presently.
Nonetheless, the detailed procedure we have followed here
shows conceptually what is involved in constructing
a geometrical RG transformation $Q$.

We note that a number of the issues regarding
locality vs. geometry arise in novel approaches 
to deconstructing gravity \cite{nima}. The RG as
a geometrical 
symmetry may prove to be a useful tool in this arena.
These and other related issues are under present
consideration.

\vspace*{1.7cm}  
\noindent  
{\bf \Large \bf Acknowledgements}  
  
I especially wish to thank A. Kronfeld for 
many useful discussions, and directing me
toward the key papers of Bell and Wilson,
and others. I also thank M. D. Schwartz for
stimulating discussions. 
Research was supported by the U.S.~Department of Energy  
Grant DE-AC02-76CHO3000.  
\frenchspacing  
  
\end{document}